\documentstyle[12pt]{article}
\newcommand{\beq}{\begin{equation}}
\newcommand{\eeq}{\end{equation}}
\newcommand{\beqa}{\begin{eqnarray}}
\newcommand{\eeqa}{\end{eqnarray}}
\newcommand{\beqar}{\begin{eqnarray*}}
\newcommand{\eeqar}{\end{eqnarray*}}
\newcommand{\tr}{{\rm tr}}
\newcommand{\ie}{{\it i.e.,}\ }
\begin{document}
\begin{titlepage}
\begin{flushright}   McGill/93-21\\
                     hep-th/9308113\\
\end{flushright}
\def\ssc{\scriptscriptstyle}
\begin{center}
   \vskip 3em
  {\LARGE Flows for rectangular matrix models }
  \vskip 1.5em
  {\large
	Ren\'e Lafrance \footnote{lafrance@hep.physics.mcgill.ca} \\
and  \\
    Robert C. Myers\footnote{rcm@hep.physics.mcgill.ca}\\[.5em] }
\em{ Department of Physics, McGill University, Montr\'eal, Qu\'ebec,
Canada H3A-2T8}
\end{center}
\vskip 1em
\begin{abstract}
Several new results on the multicritical behavior of
rectangular matrix models are presented.
We calculate the free energy in the saddle point approximation,
and show that at the triple-scaling point, the result
is the same as that derived from the recursion formulae.
In the triple-scaling limit, we obtain
the string equation and a flow equation for arbitrary multicritical
points. Parametric solutions are also examined
for the limit of almost-square matrix models.
This limit is shown to provide an  explicit
matrix model realization of the scaling equations proposed to describe
open-closed string theory.
\end{abstract}
\end{titlepage}
\newpage                           
\section{Introduction}
In recent years, the study of random matrix models as a lattice
regulator of two-dimensional euclidean gravity has shed light on the
theory of non-critical strings. The discovery of double-scaling
limit[1--3] 
has led to results to all orders in the genus expansion.
 The study of new matrix models may lead to further insights for
the continuum theory of random surfaces.

Rectangular $N \times M$ matrix models differ from previously studied
models in that they have two independent large $N$ parameters.
In general, one finds that the critical behavior is characterized by
two-dimensional singularities.
There are two special cases in these models, described by a double-scaling
limit: the vector model limit  where $M$ remains finite, and the almost-square
matrix limit where $P=N-M$ remains finite. These cases were solved
in ref.~\cite{arley}, where a surface
interpretation was also provided.\footnote{The model with $N=M$ was first
studied in ref.'s \cite{cheq,mac}.} The first limit leads to a phase of dense
polymers, and the second, to random surfaces.
Another class of critical points is described by a
triple-scaling limit, which corresponds to independently taking
both $N$ and $M$ to infinity in correlation with the approach of critical
matrix couplings\cite{polymer}.

In this paper, we present new results for the multicritical behavior of
rectangular matrix models. The remainder of this section introduces our
notation. In section~\ref{planar}, we explicitly calculate
 the free energy for the simplest non-trivial potential in the
saddle point approximation.  This calculation is performed to ensure
that the recursion formulae do not lead to spurious critical points.
 As expected, we verify that the triple-scaling calculations
correctly produce the critical behavior of the free energy.
In
section~\ref{sect_flow}, we solve the model in the triple-scaling limit,
and present the string equation for arbitrary multicritical points. The
partial differential equation arising from the potential independent
recursion relations is also formulated as a flow equation\cite{BanksDSS:90}.
Next, we examine the limit of almost-square matrices where $P$ remains
finite. We show that in this limit
these matrix models provide an explicit realization of the open-closed
string equations proposed in ref.~\cite{Dalley:b}, with $P$ playing the
r\^ole of the open string coupling constant. We briefly discuss the new
results in section \ref{conclu}.

We now review the techniques developed in ref.~\cite{polymer}
for the solution of rectangular matrix models.
Given an ensemble of $N\times M$ matrices $T$ with complex entries,
we want to study the partition function
\beq
{\cal Z}=\int dT\, \exp(-2\beta\tr V(T^\dagger T))\nonumber
\eeq
where $V(T^\dagger T)=\sum_{p=1}^{L} a_{p}(T^\dagger T)^{p}/(2p)$.
This matrix ensemble is related to the tangent space of
$U(N+M)/(U(N)\times U(M))$ \cite{arley}. One can ``diagonalize'' $T$ with
the natural $U(N) \times U(M)$ action
\beqa
T & = & V_{1} \left( \begin{array}{c}
X_{M} \\ 0	  \end{array} \right) V_{2}        \nonumber
\eeqa
where $V_{1} \in U(N)$, $V_{2} \in U(M)$ and $X_{M}={\rm diag} \left(
x_{1}, \cdots , x_{M} \right)$.
Thus the partition function reduces to
\beqa 
{\cal Z} & \propto & \int_{0}^{\infty} \prod_{i=1}^{M} \left[ dy_{i}\ \exp
	\left( -2  \beta V(y_{i}) \right)\ 
	|y_{i}|^{N-M} \right] 	\prod_{1 \leq i < j\leq M}
(y_{i}-y_{j})^{2} \label{partit}
\eeqa
where  we have set $y_{i}=x_{i}^2$.
One observes that the
partition function depends independently on $M$ and $P=N-M$.
 One could try to incorporate
the $y^{N-M}$ as a logarithmic term in the potential, but there is no
systematic way to proceed with such a potential.
Rather ref.~\cite{polymer} developed recursion
relations relating measures
$d\mu[l] \equiv dy \, y^{l} \exp(-2\beta V(y))$
with different powers $l$.
This provides a systematic framework to study all phases of these
models.\footnote{These techniques could also be applied to
Penner-like models\cite{Penner}.}

Consider
orthogonal polynomials on the positive real line
\beqa
&& \int_{0}^{\infty}\!d\mu[l] \  P_{n}^{(l)}(y)\ P_{m}^{(l)}(y)
	= \delta_{n,m}\ h_{n}^{(l)} \ \ . \nonumber
\eeqa
The recursion relations stepping in $l$ are
\beqar
	P_{n}^{(l)}(y) &=& P_{n}^{(l+1)}(y) +\phi_{n}^{(l)}
	 	P_{n-1}^{(l+1)}(y) \\*
	y P_{n}^{(l+1)}(y) &=& P_{n+1}^{(l)}(y)
	+\theta_{n+1}^{(l)}  P_{n}^{(l)}(y)\ \ \ .
\eeqar
The above coefficients also enter into the standard recursion
relation
\beqa
y P_n^{(l)} & = &  P_{n+1}^{(l)}+
\left( \theta_{n+1}^{(l)}+\phi_n^{(l)} \right)
P_n^{(l)}+\theta_n^{(l)}\phi_n^{(l)} P_{n-1}^{(l)}\ \ .
\nonumber
\eeqa
Elementary manipulations produce the potential {\it independent} relations
\beqa
	\phi_{n}^{(l)} - \phi_{n-1}^{(l+1)} =
\theta_{n}^{(l+1)} - \theta_{n}^{(l)} \quad&&\quad
	\phi_{n}^{(l-1)} \theta_{n+1}^{(l-1)}=\phi_{n}^{(l)}
		\theta_{n}^{(l)}\;\; . \label{stat_b}
\eeqa
One may also derive two potential {\it dependent} relations from
\beqa
         \frac{2n +l+1}{2 \beta} h_{n}^{(l)} &=& \int d\mu[l]\; y
        V^{\prime}(y)\  P_{n}^{(l)}(y)\ P_{n}^{(l)}(y) \nonumber \\
        \frac{n}{2 \beta} h_{n-1}^{(l+1)} &=& \int d\mu[l]\;
 	V^{\prime}(y) \left[ P_{n}^{(l)}(y) P_{n}^{(l)}(y)
	+\theta_{n}^{(l)} P_{n}^{(l)}(y) P_{n-1}^{(l)}(y) \right]\;\;.
\label{dynb}\eeqa
In the planar limit,
$\beta\rightarrow \infty$
with
$g=N/\beta$ and
$q=M/\beta$ fixed, we assume $\theta_{M\pm j}^{(P\pm i)}\to\theta$ and
$\phi_{M\pm j}^{(P\pm i)}\to\phi$~. Then, eq.~(\ref{stat_b}) is
trivial while eqs.~(\ref{dynb}) take the form
\beq
	g=\theta \partial_{\theta} U(\theta,\phi) \qquad\qquad
	q=\phi \partial_{\phi} U(\theta,\phi)          \label{dyn_planar}
\eeq
where
\[
     U(\theta,\phi) = 2 \int_{0}^{2\pi} \frac{d \lambda}{2\pi}
	 V(e^{i\lambda} +\theta +\phi +e^{-i\lambda}\theta\phi)\;\;.
\]
The critical points are identified as points where the two-dimensional
map $(g,q) \mapsto (\theta,\phi)$ is singular. Such points occur at the
vanishing of the jacobian determinant
\begin{equation}
  |J|=\left| \begin{array}{cc}
      {\partial_\theta g} & {\partial_\phi g}\\
      {\partial_\theta q} & {\partial_\phi q}
	\end{array} \right| =0\;\;. \label{jacobian}
\end{equation}
Multicritical behavior is produced by demanding higher derivatives of
linear combinations of $g$ and $q$ also vanish.
The free energy may be determined from
\[
	\theta_M^{(P)}\phi_M^{(P)} \approx
	\frac{{\cal Z}_{M+1,P} {\cal Z}_{M-1,P}}{{\cal Z}_{M,P}^{2}} \approx
	\exp(-\partial_{M}^{2} F)\;\;.
\]

\section{Saddle point approximation} \label{planar}
We begin by calculating the free energy for the
simplest non-trivial potential in the saddle point approximation
using the techniques of ref.~\cite{BIPZ}.
This calculation provides a test to verify that the triple-scaling
analysis of the recursion relations
correctly generates the critical behavior of the matrix model.
There are known cases where recursion analysis leads to spurious
results.  For example, using multiple limits (more than two) of the scaling
functions to describe eigenvalue densities on multiple intervals,
generically yields false critical points\cite{tan}.
Additional constraints producing identical minima for the potential
on each interval are required.
For rectangular matrices in the triple-scaling limit,
given that different derivations lead to
expressions for the free energy which differ by functions of $s$ (defined
below), one may suspect that the results are also spurious.
Below the nonanalytic behavior of the free energy
identified through the recursion
relations is reproduced in the
saddle point approximation for the simplest non-trivial case. This
confirms that the triple-scaling analysis of the recursion relations does
not lead to spurious results.

The planar limit of rectangular matrix
models with non-vanishing $a_{1}$ and $a_{2}$ was first solved
in ref.~\cite{cicuta}. For a general potential, the saddle point
approximation is
discussed in ref.~\cite{polymer}. We will briefly review the
results below.
The entire integrand of the partition function in eq.~(\ref{partit})
may be written $\exp (-\beta E)$ with
\beq 
    E = 2\sum_{i=1}^{M} V(y_{i})-\frac{1}{\beta}\sum_{i,j=1\atop j\neq i}^M
\ln |y_{i}-y_{j}|-\frac{N-M}{\beta} \sum_{i=1}^{M} \ln y_{i}\ \ .\label{nancy}
\eeq
Eq.~(\ref{nancy})  has an interpretation as the energy of $M$ charged
particles on a line.  The three contributions are:
an external potential, $2V(y_i)$; a Coulomb interaction
between two particles;
and an electrostatic repulsion away from the origin, with
strength $P=N-M$.
At equilibrium, the particles are confined to an
interval $[A,B]$ with  $0 \leq A \leq B$.

Following ref.~\cite{BIPZ}, the saddle point
solution of (\ref{partit}) entails the introduction of an
eigenvalue density, $\rho(z)$ which solves
\beq
     \frac{\partial V(y)}{\partial y} -\frac{P}{2\beta y}
	=\frac{M}{\beta} \int_{A}^{B}  \!\!\!\!\!\!\!\!\! - \:\:
	 \frac{dz}{y-z}\; \rho(z) \;\; .
\label{const} \eeq
One finds that $\rho(z)$ takes the form
\beq
     \rho(z) =\frac{\beta}{\pi M} u(z) \sqrt{(B-z)(z-A)}
\label{eigen_def}  \eeq
with $u(z) =\sum_{k=-1}^{L-2} h_{k} z^{k}$ where
\beq
     h_{k} = \frac{1}{2} \sum_{p=0}^{L-2-k} \frac{a_{p+k+2}}{4^p}
	 \sum_{l=0}^{p}
{p\choose l}^2(\sqrt{B}+\sqrt{A})^{2l} (\sqrt{B}-\sqrt{A})^{2p-2l}\ \ .
               \label{coeff_h}
\eeq
After integrating (\ref{const}), the saddle point free energy becomes
\beq
   F_{0} =\beta E =M\int_{A}^{B} dy \, \rho(y) \left[ 2 \beta V(y) -P \ln |y|
 \vphantom{P^P_P}  -2M \ln |y-z_{0}| \right] +\beta MV(z_{0})
	-\frac{PM}{2} \ln z_{0}
\label{free_int}  \eeq
where $z_{0}$ is a constant of integration, which does not affect the
final result.

Critical behavior corresponds to a nonanalytic dependence of the free
energy in the coupling constants in $V$. For fixed couplings, this can be
expressed as nonanalyticity in $M/\beta$ and $P/\beta$. Such behavior arises
when the eigenvalue density acquires zeroes at the
boundaries beyond those evident in (\ref{eigen_def}). This
saddle-point discussion
is connected to the planar limit description (and
eq.~(\ref{jacobian}), in particular) by the relation $|J| = 4AB u(A)u(B)$.

We are now ready to calculate the free energy for the simplest non-trivial
potential:
\beq
	V(y) = \frac{a_{1}}{2} y +\frac{a_{2}}{4} y^2\ \ .
  \label{quadratic_pot}\eeq
Using eq.'s
(\ref{eigen_def}) and (\ref{coeff_h}), the eigenvalue density is given by
\begin{equation}
	\rho(z) = \frac{1}{2\pi} \frac{\beta}{M} \left[  \left(
                a_{1} +\frac{a_{2}}{2} (A+B) \right) \frac{1}{z}  +a_2 \right]
	\sqrt {(B-z)(z-A)}\ \ .
\label{eigen} \end{equation}
We evaluate (\ref{free_int}), using (\ref{eigen}) and setting
$z_{0}=(A+B)/2$, to obtain
\begin{eqnarray}
\frac{F_{0}}{\beta^2} &=& \frac{(B-A)^2}{32} \left[ {a_{1}}^2
	+\frac{5}{4} a_{1}a_{2} (A+B) + \frac{a_{2}^{2}}{32}
	 (9A^2 +9B^2 +14AB ) \right]  \nonumber \\
& & +\frac{P}{8\beta} \left[  a_{1}\left(\sqrt{B}-\sqrt{A}\right)^2
       +\frac{a_{2}}{8}\left( 3B^2 +3A^2 +2 AB -4(A+B)\sqrt{AB} \right)
 	 \right]  \nonumber  \\
& &  +\frac{M}{4\beta} \left[ \frac{a_{1}}{2} (A+B)
              +\frac{2M}{\beta}+\frac{P}{\beta}\right]
	- \frac{P}{\beta} \left[ \frac{2M}{\beta} +\frac{P}{\beta}
	 \right] \log \frac{\sqrt{B}+\sqrt{A}}{2} \nonumber  \\
& & -\left( \frac{M}{\beta} \right)^2   \log \frac{B-A}{4}
	+\left( \frac{P}{2\beta} \right)^{2} \log AB\ \ . \label{free_before}
\end{eqnarray}
The endpoints $A$ and $B$ are determined by
\begin{eqnarray}
	\frac{2M}{\beta} +\frac{P}{\beta} &=& \frac{a_{1}}{2}
	(A+B) +\frac{a_{2}}{8} (3A^2+3B^2 +2AB) \\
	\frac{P}{\beta}  &=& \sqrt{AB} \left( a_{1}
	 +\frac{a_{2}}{2} (A+B)  \right)\ \ .
\end{eqnarray}
Given these expressions, it is not at all apparent which matrix potentials
will yield nonanalytic behavior in the free energy. Our approach is
to determine the critical point from the planar string equations
(\ref{dyn_planar}), and then examine the free energy (\ref{free_before})
for nonanalytic behavior using the relations\cite{polymer}
\beq
A=(\sqrt{\theta}-\sqrt{\phi})^2\qquad\qquad
B=(\sqrt{\theta}+\sqrt{\phi})^2\ \ .
\label{boundaries}
\eeq

For the potential (\ref{quadratic_pot}),
the vanishing of the jacobian determinant becomes
\beq
    |J|=a_{1}^{2}+4a_{1}a_{2}(\theta_{c}+\phi_{c})
	+4a_{2}^{2}(\theta_{c}^2 +\theta_{c}\phi_{c}
	+\phi_{c}^{2})=0
\label{vanished}
\eeq
Setting $g_{c}=1=\theta_{c}$ and $\phi_{c}=y^2$ with $0 \leq y\leq 1$,
eq.'s (\ref{dyn_planar},\ref{vanished}) yield
\beq
	a_{1}=2 \frac{1+y+y^2}{2y+1} \qquad
	a_{2}=-\frac{1}{2y+1} \qquad
        q_{c} = \frac{y^{3} (2+y)}{(2y+1)}
\label{critical_value}
\eeq
One easily confirms that for these couplings the eigenvalue density
$\rho(z)$ has an extra zero at $z=B=(1+y)^2$.

To study the singularity, one expands (\ref{dyn_planar}) around these
critical values
\beqa
         \Delta g &=& -\frac{2}{2y+1} (\Delta \phi -y\Delta \theta)
	-\frac{1}{2y+1} (\Delta \theta^{2}
	+2\Delta \theta \Delta \phi) \nonumber\\
         \Delta q &=&\frac{2y}{2y+1} (\Delta \phi -y \Delta \theta)
       -\frac{1}{2y+1} (\Delta \phi^2 +2 \Delta \theta \Delta \phi)
  \nonumber
\eeqa
where $\Delta$ is used to denote $\Delta x =x-x_{c}$.
One observes that both $\Delta g$ and $\Delta q$ are
proportional to $\Delta \hat{\phi}=\Delta \phi -y\Delta \theta$, and that
$\Delta \hat{q}=\Delta q +y\Delta g$ has no linear variation.
Further for generic
rectangular matrices (\ie $0 <y<1$), it is convenient to define
\beqar
       \hat{g} = g-\frac{2y+1}{3y(y+1)} \hat{q} &\qquad&
       \hat{\theta} = \theta +\frac{2y+1}{3y(y+1)} \hat{\phi}\ \ .
\eeqar
With these choices, the planar equations become
\beqa
      \Delta \hat{g} &=& -\frac{2}{2y+1} \Delta \hat{\phi}
         +\frac{2(y^2+y+1)}{3y(2y+1)(y+1)}\Delta \hat{\theta}\Delta \hat{\phi}
         +\frac{(y-1)(y+2)}{9y^2(y+1)^2}\Delta \hat{\phi}^2\nonumber \\
      \Delta \hat{q} &=& -\frac{1}{2y+1} \left[
	3y(y+1) \Delta \hat{\theta}^{2}
	-\frac{y^{2}+y+1}{3y(y+1)} \Delta \hat{\phi}^{2} \right]\label{qhat}
\eeqa
The quadratic form of the singularity in (\ref{qhat}) suggests
$\Delta \hat{q} \approx -\delta ^{2} t$, where $\delta=\beta^{-2/5}$
just as for quadratic singularities in double-scaling[1--3].
Then eq.~(\ref{qhat}) yields
 $\Delta \hat{g} =-\delta s$, $\Delta\hat{\theta}=-\sum_p \delta^p f_p$
and $\Delta\hat{\phi}=-\sum_p\delta^p h_p$, where $f_p$ and $h_p$ are solved
for in terms of $s$ and $t$.

Upon inserting these scalings in eq.~(\ref{free_before}), the
planar free energy becomes
\beq
     F_{0} =  \frac{12y(y+1)}{5(2y+1)^{2}}
	\left[ \frac{(y^2+y+1)(2y+1)^{2}}{36y^{2}(y+1)^{2}}
s^{2} +\frac{2y+1}{3y(y+1)} t
	 \right] ^{5/2}\ \ . \label{end_result}
\eeq
This expression differs
from the planar contribution to free energy presented for this model in
ref.~\cite{polymer}, only by rescalings of $s$
and $t$, which were made there (and also in the next section)
to simplify the results.
Without such rescalings, both expressions would be identical.
This demonstrates that the recursion relations successfully determine
the correct critical behavior for the matrix model.
Note that non-universal contributions to $F_0$ were dropped from
({\ref{end_result}). In fact, some of these terms are actually
divergent, being proportional to $\beta^{\alpha/5}$ with $\alpha >0$, but
they are analytic in both $s$ and $t$.
Similar divergent analytic terms appear in the scaling analysis of
the saddle point approximation for the hermitian matrix model\cite{mylaf}.
A finite term analytic in $s$ was also dropped. In general, one might
expect additional finite terms which are nonanalytic in $s$\cite{polymer},
but they do not arise for this critical point.

\section{Triple-scaling limit} \label{sect_flow}

In this section, we reexamine the triple-scaling anaylsis of
ref.~\cite{polymer}. We formulate the potential independent relations
in terms of flow
equations, and this allows us to determine the general string equation
for this class of multicritical points.

In these models since there are two parameters $N$ and $M$ which
diverge separately as $\beta\rightarrow\infty$, it is natural to
begin with a general scaling ansatz for two linearly independent
combinations of $P$ and $M$
\[
\frac{AM+CP}{\beta} = \left[  \frac{AM+CP}{\beta} \right]_{c}
	 -\beta^{\nu -1} t \qquad\quad
\frac{BM+DP}{\beta} = \left[ \frac{BM+DP}{\beta} \right]_{c}
	 -\beta^{\hat{\nu}-1} s\ \ .
\]
In the scaling limit then, finite differences become derivatives:
$\partial_{M} = -\beta^{\nu} A \partial_{t}
	-\beta^{-\hat{\nu}}B \partial_{s}$, and
$\partial_{P} = -\beta^{-\nu}C\partial_{t}
	-\beta^{-\hat{\nu}}D \partial_{s}$.
The planar limit analysis suggests $\hat{\nu}=3\nu$.
As in the previous section, we set
$\theta_{c}=1$ and $\phi_{c}=y^{2}$, and define $\delta=\beta^{-\nu}$.
We then introduce
\beqa
   \theta_{M+n}^{(P+l)} &=& 1-\exp \left[
       -\delta (An+Cl) \partial_{t} -\delta^{3} (Bn+Dl) \partial_{s}
       \right] \sum_{q=2}^\infty \delta^{q} h_{q}(t,s)
 \nonumber \\
    \phi_{M+n}^{(P+l)} &=& y^{2} -\exp \left[
         -\delta (An+Cl) \partial_{t} -\delta^{3} (Bn+Dl) \partial_{s}
          \right] \sum_{q=2}^\infty \delta^{q}k_{q}(t,s)
 \label{ansatz_p}
\eeqa

First we determine the scaling limit of eq.~(\ref{stat_b}), which are
independent of the matrix potential. Inserting
(\ref{ansatz_p}), leads to $k_{2}(t,s) = y ( h_{2}(t,s) +g_{2}(s) )$ and
a partial differential equation
\beq
   \partial_{s} h_{2} = \frac{C^{2}}{4T} \frac{y-1}{y^2} \partial_{t}
	\left[ h_{2}^{2} -\frac{C^{2}}{6} \frac{y+1}{y}
	 \partial_{t}^{2} h_{2} \right]
	-\frac{C^{2}}{2Ty^{2}} g_{2}\partial_{t} h_{2}
	-\frac{1}{2} \partial_{s} g_{2}
\label{flow_before}  \eeq
where $T=AD-BC$. It was noted in ref.~\cite{polymer} that
neglecting the terms involving $g_{2}$, eq.~(\ref{flow_before})
is the KdV equation with $s$ and $t$ playing the
r\^oles of the time and space coordinates, respectively.
Presently, we also point out that eq.~(\ref{flow_before}) has the
form of a flow equation. This connection is made clear by first
introducing $h(t,s)=h_2(t,s)+g_2(s)/2$, and then using (some of) the
freedom to rescale
$s \rightarrow \alpha s$, $t\rightarrow \beta t$, $h \rightarrow
\gamma h$, and $g_{2}\rightarrow \eta g_{2}$. Thus eq.~(\ref{flow_before})
can be expressed as
\beq
	\partial_{s} h = \partial_{t}\left(  2g_{2}(s)\,R_{1}[h]
               + R_{2}[h] \right)
\label{flow_after}\eeq
where $R_{1}[h]=-h/4$ and $R_{2}[h]=(3h^2-\partial_t^2h)/16$  are the first
and second Gel'fand-Dikii differential polynomials\cite{GD_poly}.

Eq.~(\ref{flow_after}) may be compared to the flow equations which
are usually discussed in the context of matrix models and the KdV
hierarchy\cite{BanksDSS:90}. Given the string equation for the general
model interpolating between multicritical points
\beq
t=\sum_{k=0}^\infty (k+\frac{1}{2})\mu_k R_k[h]\ \ ,
\label{general}
\eeq
the generalized KdV equations arise as flow equations for
$h(\{\mu_k\},t)$
\beq
\frac{\partial h}{\partial\mu_k}=\partial_t R_{k+1}[h]\ \ .
\label{float}
\eeq
If the first two couplings in the string equation (\ref{general})
were correlated as $\mu_1=s$ and $\frac{\partial\mu_0}{\partial s}=2g_2(s)$,
eq.~(\ref{float}) would lead to a flow equation of the form given in
eq.~(\ref{flow_after}). We reiterate that these results follow without
restricting the matrix potential in any way. The only assumption is
to fix the ratio of the scaling exponents, $\hat\nu/\nu=3$ (and
$q\ge2$). We have not found any other choices of this ratio which
lead to interesting results.

Next we would like to examine the string equations which arise from the
potential dependent relations (\ref{dynb}). For a quadratic potential
(\ref{quadratic_pot}) with the critical values (\ref{critical_value})
and $\nu=1/5$, one finds $g_2=s/2$ and
\beq
t=\frac{s^2}{4} R_{0}[h] +\frac{3}{2}s R_{1}[h]
-\frac{9}{2}\left(\frac{y+1}{y-1} \right)^{2}  R_{2}[h]
\label{one_eq}
\eeq
where $R_0[h]=1/2$ and we have used the remaining freedom in rescaling
variables to simplify these results.\footnote{This string equation
(\ref{one_eq}) is the same as that presented in ref.~\cite{polymer}, but
with slightly different rescalings, and a distinct combination of $h_2$
and $g_2$ for the scaling function.}
Comparing eq.'s (\ref{one_eq}) and (\ref{general}),
one sees that $\mu_0$ and $\mu_1$ have
precisely the dependence on $s$ to be compatible with the flow equation
(\ref{flow_after}).

Ref.~\cite{polymer} also presents results for the critically tuned
quartic potential which produces a third order multicritical point.
Recasting those results in the form of eq.~(\ref{general})
by a rescaling, one finds that $g_2=(s/6)^{1/2}$ and
\beq
t=4\left(\frac{s}{6}\right)^{\frac{3}{2}} R_{0}[h] +\frac{3}{2}s\ R_{1}[h]
-\frac{27}{2}\left(\frac{y+1}{y-1} \right)^{4}  R_{3}[h]
\label{two_eq}
\eeq
where $R_{3}[h]=-(10h^3-10h\partial_t^2h-5(\partial_th)^2+\partial_t^4h)/64$.
We have also carried out the triple scaling analysis for the $k=4$
multicritical point, which is produced by tuning a sextic potential.
The results are similar to those above with $g_2\propto s^{1/3}$ and
in eq.~(\ref{two_eq}), $s^{3/2}\rightarrow s^{4/3}$ and $R_{3}[h]\rightarrow
R_{4}[h]$, as well as changing numerical factors. For higher
order critical points, we found it impractical to explicitly carry out the
complete triple-scaling analysis, but the form of the general string
equation is clear from the above examples. It takes the form of
eq.~(\ref{general}) with the only nonvanishing coefficients being
$\mu_0$, $\mu_1$ and $\mu_k$. The values of $\mu_0$ and $\mu_1$ are
fixed by the analysis of the potential independent
equations (\ref{flow_after}). So it
remains only to fix $g_2$ and $\mu_k$. The full triple-scaling analysis
is unnecessary to determine these coefficients, but rather they can
be extracted from the planar limit alone. At the ($k$+1)'th multicritical
point, we find
\[   g_{2}  =\left[  \frac{(k-1)! k!}{2(2k-1)!} \ s \right]^{1/k} \]
and the string equation becomes
\beq
       t=\frac{k\,sg_2}{k+1} R_{0}[h]\ +\
	\frac{3}{2} s\, R_{1}[h]\ -\
	\frac{3^{k+1}}{2} \left( \frac{y+1}{y-1} \right)^{2k}R_{k+1}[h]\ \ .
\nonumber \eeq
The string equation is that of the conventional ($k$+1)'th order multicritical
point but perturbed by the operators $R_{1}$ and $R_{0}$. The new scaling
parameter $s$ governs the strength of the perturbations.

\section{Parametric solutions}

A seperate class of novel critical points is governed by coupled
differential and finite difference equations. These occur in the
special limits where only a single large $N$ parameter diverges.
At present, we have no further remarks on the case of the
vector models. We wish to point out the connection
of the almost square matrix models, in which $P=N-M$ remains a finite
parameter, to the open--closed string equations proposed by
ref.~\cite{Dalley:b}.
The simplest critical point was discussed in ref.~\cite{polymer}
where the scaling function satisfies the Painlev\'e II equation with
a constant
\[
\frac{1}{2}\partial^2_sk^{(P)}_1-k^{(P)3}_1+sk^{(P)}_1=P+\frac{1}{2}
\]
($k^{(P)}_1$ and $s$ will be defined below). Comparing to
ref.~\cite{Dalley:b} suggests that $P$ plays the
role of an open string coupling constant. In these parametric
models, the potential independent recursion relations give rise
to a finite difference relation
\beq
k^{(P)2}_1-\partial_sk^{(P)}_1=k^{(P+1)2}_1+\partial_sk^{(P+1)}_1\ \ .
\label{finite}
\eeq
The same relation was also developed for scaling functions satisfying
a string equation in the mKdV hierarchy with coupling constants which
differ by one\cite{Dalley:b}. To confirm
the connection of these parametric critical
points to open--closed strings, we consider the next multicritical
point.

This multicritical point can be produced by tuning a cubic potential,
$V(y) = \frac{a_{1}}{2}y  +\frac{a_{2}}{4}y^{2}+\frac{a_{3}}{6}y^{3}$.
Square matrices correspond to $g_{c}=q_{c}=1$, where the latter is a choice
of normalization as is $\theta_{c}=1$.
These multicritical points are related to approach of zeroes in the
eigenvalue density to the origin. So first we choose $\phi_c=1$, which
by eq.~(\ref{boundaries}) yields $A=0$ and $B=4$.
Now eq.'s (\ref{dyn_planar}) and (\ref{jacobian}) fix
$a_{1} = -2+2a_{3}$ and $a_{2} = 1-4a_{3}$.
Then, the eigenvalue density is given by
\[
       \rho(z) =\frac{\beta}{\pi M} \sqrt{(4-z)z} \left( 1-2a_{3}
	 + a_{3}z \right)\ \ .
\]
Choosing $a_3=1/2$ yields the form
$\rho(z) \propto z^{3/2}$ at the origin.

For a scaling solution, we define $\delta \equiv \beta^{-\nu}$ and we
use the following scaling ansatz: $\frac{M}{\beta} =
1-\delta^{\gamma}s$ and
\beqa
         \theta_{M+n}^{(l)} + \phi_{M+N}^{(l)} &=& 2
	-2\exp \left( -n\delta \partial_{s} \right) \sum_{p=2}
	 \delta^{p} h_{p}^{(l)}(s)  \nonumber  \\
         \theta_{M+n}^{(l)} - \phi_{M+N}^{(l)} &=&
	-2\exp \left( -n\delta \partial_{s} \right) \sum_{p=1}
	 \delta^{p} k_{p}^{(l)}(s)   \label{ansatz_b}
\eeqa
Derivatives do not replace finite differences in $l$ because $P$ does
not scale. Given eq.~(\ref{ansatz_b}) without any further assumptions,
the potential independent relations
(\ref{stat_b}) yield eq.~(\ref{finite}) which relates the scaling
function at different values of $P$.
The identical equation is found in ref.~\cite{Dalley:b}
to relate solutions of the
unitary matrix models with different numbers of flavours of
`quarks'\cite{joe}.
The scaling of the potential dependent relations (\ref{dynb}) yields
$\nu=1/5$, $\gamma=4$,
\beq
      h_{2}^{(P)}(s) =\frac{\partial_{s}k_{1}^{(P)} -k_{1}^{(P)2}}{4}
\nonumber
\eeq
and the scaling function $k_{1}^{(P)}$ satisfies
\beq
      s k_{1}^{(P)} -k_{1}^{(P)5}
	+\frac{5}{3} k_{1}^{(P)} (\partial_{s} k_{1}^{(P)})^{2}
	+\frac{5}{3}k_{1}^{(P)2} \partial_{s}^{2} k_{1}^{(P)}
	-\frac{1}{6} \partial_{s}^{4}k_{1}^{(P)} =P+\frac{1}{2}
\label{mkdv2}  \eeq
where we have shifted $k_{1}^{(P)} \rightarrow \alpha k_{1}^{(P)}$ and
$s \rightarrow s/\alpha$ with $\alpha = [8/3]^{1/5}$. This is
the second equation in the mKdV hierarchy, where again $P$ appears as the
open string coupling in the constant term.

Using eq.~(\ref{ansatz_b}), one obtains the free energy
\beq
	\partial_{s}^{2} F^{(P)} =-\frac{1}{2} (k_{1}^{(P)2}
	+\partial_{s} k_{1}^{(P)} )
\label{fizzx}
\eeq
Here, the specific heat is proportional to the Miura map of the scaling
function. Defining $u \equiv k_{1}^{(P)2} +\partial_{s}k_{1}^{(P)}$,
using eq.~(\ref{mkdv2}) may express the string equation in terms of the
specific heat
\beq
        u R^{2}[u] -\frac{1}{2}R[u]R^{\prime \prime}[u]
	+\frac{1}{4} (R^{\prime}[u])^{2} =P^{2}
\label{RGE}   \eeq
where $R=R_{2}[u]-s$.
This is the original form of the scaling equations proposed for open-closed
strings with the open string coupling equal to $P$\cite{Dalley:b,Dalley:a}.
For $P=0$, the perturbative solution of eq.~(\ref{RGE}) is simply
$R[u]=0$, which in the present case is the Painlev\'e I equation.
As well though, eq.~(\ref{RGE}) provides a nonperturbative solution
for two-dimensional gravity\cite{Dalley:a}. The rectangular matrix
models provide an explicit matrix model realization of these scaling
equations extended to a nonvanishing open string coupling constant.

\section{Discussion} \label{conclu}
Rectangular matrix models display a rich variety of multicritical
behaviors. Underlying this diversity is the fact that these models
have two independent large $N$ parameters, which leads to multicritical
behavior governed by two-dimensional singularities.

Based on the analysis of the recursion relations alone, the connection
of various critical points to singular behavior of the actual
matrix model is not always clear. For the triple-scaling points,
we have verified the validity of critical behavior by recovering the same
nonanalytic planar contribution from a saddle point analysis. The
multicritical points for the triple-scaling limit are governed by
string equations which are simply expressed in terms of the Gel'fand-Dikii
differential polynomials\cite{GD_poly}. The usual $k$'th order string equation
in the KdV hierarchy is perturbed by the $R_1$ and $R_0$ operators.
The corresponding coupling $\mu_1$ and $\mu_0$ are functions of the
new scaling parameter, and a flow equation expressing this dependence
naturally arises.
The effect of the $R_0$ can be absorbed in a renormalization of the
cosmological constant to $\tau = t+ X s^{k/(k-1)}$, where $X$ is some
numerical constant. For small $s$, as well as the usual genus
expansion in $\tau^{-(2k+1)/k}$, there is also an expansion
in $s/\tau^{(k-1)/k}$ at each genus due to the perturbation of $R_1$
in the string equation. In ref.~\cite{polymer}, it was conjectured that
the new dynamics uncovered by the triple scaling analysis should be related
to a gas of punctures arising from the distinction between $M$ and $N$ loops
in the surface interpretation\cite{arley}.
In the present context then, it appears that tuning
for triple scaling involves tuning these punctures to behave
as a linear combination of the $R_0$ and $R_1$ operators. One might expect
that a more subtle scaling would lead to more complex perturbations
by linear combinations involving higher order Gel'fand-Dikii polynomials,
but as yet we have been unsuccessful in producing such a tuning.

The present reformulation of the triple scaling results provides some
insight into the interpretation of the dual expansion for large $|s|$,
which was noted in ref.~\cite{polymer}. The $R_1$ perturbation
expansion in $|s|/\tau^{(k-1)/k}$
is expected to have a finite radius of convergence, and beyond that point
the solution should be expanded in terms of $\tau/|s|^{k/(k-1)}$.\cite{moore}
In this domain, the model is expected to be in the neighbourhood
of the ``topological'' model governed by $R_1[h]$. (This crossover
behavior can be explicitly seen for the $k=2$ and 3 critical points, at least
in the planar limit where the string equations are easily solved.)
For this interpretation to be applicable, one should consider the real
root of the planar equation which vanishes as $|s|\rightarrow\infty$.
This leads to the expected
expansion, $h\approx-\frac{8}{3}\frac{\tau}{s}+\ldots$. Note that in this
phase, the full expansion is not entirely a perturbation expansion in
$\tau/|s|^{k/(k-1)}$ which is the planar contribution of $R_k[h]$,
but rather the derivative terms in $R_k[h]$ produce higher genus contributions
as well. This is not the complete story though.
For large $|s|$, the planar equation may have another real root which
yields an expansion $h\propto s^{1/(k-1)}+\ldots$ (here if ($k$--1) is even,
one requires $s>0$,  and then there is a second root at $-s^{1/(k-1)}$).
These were
the dual expansions presented in ref.~\cite{polymer}. In these
cases, both $R_1$ and $R_k$ are equally important in determining
the behavior of the perturbative expansion. The interpretation of these
extra strongly coupled phases remains unclear. Similar strongly coupled
domains may also separate the $R_1$ and $R_k$ phases, if there is a gap
between the regions of convergence of the expansions in $|s|/\tau^{(k-1)/k}$
and $\tau/|s|^{k/(k-1)}$.

Finally, we have examined the parametric solutions for
almost-square matrices. By examining the $k=2$ critical point, we
confirmed that these solutions provide an explicit matrix model
realization of  the scaling equations for open-closed strings\cite{Dalley:b},
with $P=N-M$ playing the r\^ole of the open string coupling constant.
This is a natural position for $P$ to appear in, since in the limit
of almost-square matrices, the surface interpretation associates $P$ with
introducing punctures or boundaries\cite{arley}.
The scaling equations, which naturally arise with our scaling ansatz
(\ref{ansatz_b}), are those of the mKdV hierarchy but with a constant
$P+\frac{1}{2}$.
The same equations naturally arise in the double-scaling limit of
unitary matrices coupled to $C=P+\frac{1}{2}$ flavors of quarks\cite{joe}.
We emphasize though that the physics of these models is not identical.
In the present case, the free energy is simply related to the Miura
map of the scaling function (\ref{fizzx}). For the unitary matrix
models, one has simply $\partial_{s}^{2} F^{(P)} =-\frac{1}{2} k(s)^2$,
where $k(s)$ is the corresponding scaling function\cite{joe}.
Thus even though one has a simple map between solutions in the KdV and mKdV
hierarchies\cite{Dalley:b}, the physics of unitary and rectangular matrix
models remains distinct.

\section*{Acknowledgments}
This research was supported by NSERC of Canada and Fonds FCAR du
Qu\'ebec. R.C.M. would like to thank Vipul Periwal for useful conversations.

\end{document}